\documentclass[prl,showpacs,preprintnumbers,amsmath,amssymb]{revtex4}
\usepackage{graphicx}
\usepackage{dcolumn}
\usepackage{bm}
\begin{document}
\title{Elastic scattering at 7 TeV and high energy cross section for cosmic ray studies}
\author{T. Wibig}

\affiliation{Physics Dept., University of \L \'{o}d\'{z};
National Centre for Nuclear Research, Cosmic Ray Laboratory, \L od\'{z},
Uniwersytecka 5, 90-950 \L \'{o}d\'{z}, Poland}

\begin{abstract}
The recent measurements of the elastic cross section by the TOTEM Collaboration 
together with the first estimations of the inelastic cross 
sections by other LHC detectors 
are used to test the simplest version of the geometrical model of the proton-proton scattering. 
We show that the description found for lower energy data, with the modest adjustment of the
model parameter extrapolation,
could be, in principle,  
used to describe the LHC measurement and to predict the cross sections in very high
energy cosmic ray domain. 
However, the shape of the first elastic
dip in the elastic differential scattering cross section 
suggests that ratio of the real to the imaginary 
part of the elastic amplitude is falling rather fast and 
the analysis of the elastic cross section fraction suggests that the 
geometrical picture of the proton-proton collision should be
modified  considerably 
when entering the ultra high-energy domain.
\end{abstract}


\maketitle

\section {Introduction}
The cross section of a very high energy proton (and nuclei) interaction is 
one of the most significant properties of the multiparticle production process 
determining development of the cascade emerging when extremely 
high energy astroparticle enters the Earth atmosphere. The phenomenon 
called Extensive Air Shower (EAS) is the one and only trace leading 
to the nature and origin of these particles. The question of the cross sections
is of great 
importance since the mid 60s when the event with estimated total 
energy exceeding $10^{20}$ eV was detected by Linsley \cite{linsley} and when 
later the 
limit called now "GZK cut-off" has been announced \cite{gzk}. The partial solution 
of the ultra high-energy cosmic ray puzzle is the proposition of the heavy primaries. 
It moves the limit of the GZK mechanism to the limit defined by the 
photo-nuclear disintegration 
via the giant dipole resonance and the acceleration effectiveness by the 
factor of $Z$ (26 in the case of iron). Obvious consequence of the change of 
the primary particle from the single nucleon to the nucleus, which is a cluster of $A$ 
nucleons (56 in the case of iron), 
is the change of the picture of the emerging 
EAS cascade. The "0-th  moment" of the particle distribution along the 
shower path is still proportional to $E$, the total energy of incoming 
particle, but "the first moment"  describing "average" particle position 
in the shower (measured for years in the form of $X_{\rm max}$ -  the 
position of the maximum of the shower development) depends on $E/A$, the energy 
per nucleon and, of course, on the interaction cross section. 
This shows how important 
the detail knowledge of the energy dependence of the interaction cross 
section is. 
The highest energy data available at the beginning of the new 
millennium came from the experiments at FNAL at $\sqrt{s}$  
about 2 TeV. It is almost 2 orders of magnitude 
below the observed highest energy cosmic ray events (in the centre of mass system (c.m.s.), 
in the laboratory frame of references the difference is of 4 orders). 
All what could be done
for ultra-high energy cosmic ray studies 
was to extrapolate the Tevatron, SPS, ISR and other low energy results in a 
``reasonable'' way. The example of such extrapolations with related uncertainties 
is discussed in Ref.~\cite{ulrich}.

Some theoretical suggestions can be directly  helpful as, e.g., Froissart limit 
which gives the asymptotic binomial in $\ln{(s)}$ form.
There exists also the concept of the description of the 
scattering of extended objects on each other based on the optical analogy. 
The complex opacity functions convolution leads 
to the elastic scattering amplitude, and than, through the optical theorem, 
to the total collision cross section and the inelastic and elastic cross sections.
The only unknown is the 
``hadronic matter distribution'' of colliding hadrons at the particular c.m.s. available energy.
Many propositions can be found in the 
literature. The degree of sophistication depends on what, and to what extend, 
is supposed to be described. Some examples are discussed in Ref.~\cite{menon93,menon,achilli}.  
Because we are interested mainly in the ``reasonable'' high-energy 
inelastic cross section extrapolation, details of the differential 
elastic scattering are not decisive for us as, e.g., the very high momentum transfer amplitudes. However, the word ``reasonable'' used 
above means that we should try to find the description which is not in a
strong disagreement with all what is known and measured. We would like to pay a special attention 
to the feature of the first diffractiv dip in the elastic differential cross section measured 
recently at 7 TeV by the TOTEM Collaboration \cite{totem-dif}.

We are going 
to try the simplest solution which, as was shown in Ref.~\cite{twds}, works well 
up to TeV energies, and to see if this description should be modified when
applied for the available at present LHC data
\cite{totem-dif,lhc-totem,cms,alice,atlas}.
It is obvious that the studies of the hadron shape are expected to
lead to the kind of scaling (of the opacities) as the interaction energy 
changes. The two simples solutions: geometrical scaling and the 
factorisation hypothesis separately do not give the satisfactory 
descriptions. We will check if the modified geometrical scaling
proposed in Ref.~\cite{twds} works up to LHC energies.

The scaling properties determine the
rise of the proton-proton cross section (total, inelastic) as the
interaction energy increases, what, as it was said, is an important feature of the strong
interaction picture from the EAS perspective. 
The rise itself is established quite well both
from theoretical and experimental point of view. However, the question
how fast does the cross sections rise is discussed permanently and 
the definite answer
is still lacking. Theoretical predictions agree well with one another and
with accelerator data in the region where data existed ($\sqrt{s} \sim 20
\div 2000$ GeV) but they differ above. Before the LHC the only information
can be derived from the cosmic ray extensive air shower (EAS) data.
The important difference between the collider and EAS arrays proton--proton
cross section measurements is that in the EAS development
the proton--air interactions are involved. The value which is really measured 
is the cross section for the interactions with air nuclei. 
The value of the proton--proton cross section is obtained
from this kind of data using the theory of nuclei interactions, e.g., the Glauber model
\cite{glauber}, and/or 
extensive Monte-Carlo calculations.

Two large cosmic ray EAS experiments, Akeno \cite{akeno} and Fly's Eye \cite{FE}, gave
estimations of the proton-Air cross section at about $\sqrt{s}
\approx 10^4$ GeV in the previous millennium. Next two giant hybrid: surface 
and air-fluorescence, arrays: High Resolution Fly's Eye (HiRes) in Utah
 and Pierre Auger Observatory (PAO) in Argentina have reported \cite{hires,pao} 
new data of the
cascade development at very high energies and the estimated 
 proton-Air cross sections above the primary cosmic ray particle energy of $10^{18}$~eV.
We will use this data for comparison.

\section {Modified geometrical scaling model}
Introducing the impact parameter formalism cross sections can be
described using one, in general complex, function $\chi$
in the form
\begin{eqnarray}
{\sigma}_{\rm tot}~=~2\:\int\:\left[\:1\:-\:{\rm Re}\left (
{\rm e}^{i \chi ({b})}\right)\: \right]
d^2 {\bf b}~~,
\nonumber \\
{\sigma}_{\rm el}~=~\int\:| \:1\:-\:{\rm e}^{i \chi ({b})}\:| ^2\:
d^2 {\bf b}~~,
\label{sigsig}
\nonumber \\
{\sigma}_{\rm inel}~=~\int\:1\:-\:| \:{\rm e}^{i \chi ({b})}\:| ^2\:
d^2 {\bf b} ~~.
\label{xsecs}
\end{eqnarray}

The phase shift $\chi$ is related to the scattering amplitude
by the two dimensional Fourier transform

\begin{eqnarray}
1\:-\:{\rm e}^{i\chi({\bf b})}
~=~{\frac {1}  {2\: \pi \: i}}\int\:{\rm e}
^{-i{\bf b\:t}} S({\bf t}) d^2 {\bf t} ; \nonumber \\
S({t}) ~=~{\frac i  {2\: \pi \: }}\int\:{\rm e}
^{ i{\bf b\:t}} \left(
1\:-\:{\rm e}^{i\chi({\bf b})} \right)
d^2 {\bf b} .
\label{eq2}
\end{eqnarray}

With the optical analogy one can interpret $
1\:-\:{\rm e}^{i\chi({\bf b})}
$ as a transmission coefficient
for a given impact parameter. Considering two colliding object we
assumed 

\begin{equation}
\chi({b},s)~=~i\: \omega(b,s)~=~i\: F\int \: d^2{\bf b'}\:
\rho_a({\bf b})\rho_b({\bf b}\:+\:{\bf b'}) .
\label{rho}
\end{equation}

\noindent
where $\rho_h$ is a particle "opaqueness"
(the matter density integrated
along the collision axis). The optical interpretation is
supplied by a dependence on the
interaction energy, $s$. 

The phase shift $\chi$ is, in general, 
a two-variable complex function. 
There are two ways of making this description more attractive:
the factorisation hypothesis (FH) and geometrical scaling (GS):
\begin{eqnarray}
\chi(s,\:b) ~=~ i\: \omega(b)\:f_F(s) ~~~~~~{\rm (FH)} \nonumber \\
\chi(s,\:b) ~=~ i\: \omega (b\: \times f_{GS}(s)) ~~~~~~{\rm (GS)}
\label{gsfh}
\end{eqnarray}

From the optical point of view, the FH means that the hadron is
getting blacker as the energy increase, while the GS means that it is getting
bigger and functions $f_F(s)$ and $f_{GS}(s)$ describe the energy behaviour of this 
proton ``changes''. Analysis of the elastic data above $\sqrt{s} \sim 20$ GeV
shows that none of the propositions given in Eq.(\ref{gsfh})  is realised
exactly
(\cite{chouyang,ISR}).
 
The main information about the hadron phase shift function $\chi$ comes from
elastic scattering experiments, more precisely: from the measured differential
elastic cross section. In the present work the form of the phase shift which 
follows the original GS idea \cite{buddd} is adopted. It differs from the well-known 
Martin's formula where the ratio of
the real to the imaginary part of the scattering amplitude depends also on the
momentum transfer. We have assumed, after \cite{men1}

\begin{equation}
 \chi(s,b)  ~=~(\lambda(s)+i)\: \omega(b,s)~~,
\label{lambdach}
\end{equation}

The accurate enough data description has been found \cite{ws1} with
the simplest and quite old 
\cite{heis}
proposition of the
``matter'' distribution in the form

\begin{equation}
\rho_h({\bf b})~=~\int d z {\frac { m_h}  {8 \pi}} {\rm e}^{-m_h {\bf r}}
\label{pdens}
\end{equation}

\begin{figure}
\centerline{    
\includegraphics[width=10cm]{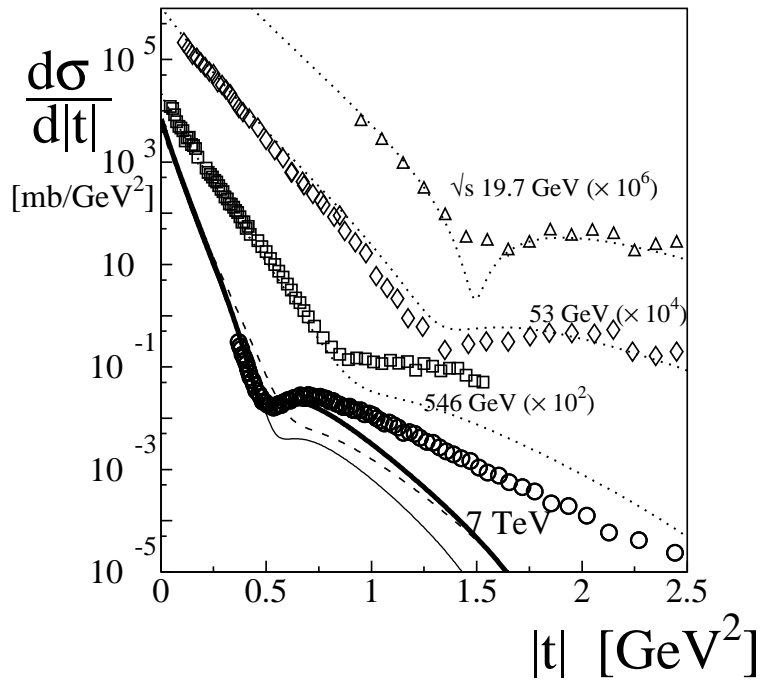}}
\caption{Differential elastic cross section for different energies. 
Low energy data from FNAL, ISR and SPS \cite{datalowel} are shown by triangles, diamonds 
and squares, respectively. The dotted lines calculated for respective lower energies 
and the dashed line for 7 TeV
are obtained with the original parameters of Ref.~\cite{twds}.
The thin solid line is obtained with the adjustment of $\lambda$ and $f_{GS}$ parameters only,
while the thick solid line with additional the $f_F$ parameter adjustment (see text).}
\label{gscal}
\end{figure}

The quality of the proposed parametrisation is presented in Fig.\ref{gscal}.

\section{Results}
Following the modified geometrical scaling fits found in 
Ref.~\cite{twds} we have used for $\sqrt{s} = 7$~TeV values of $\lambda = 0.16$ and 
$f_{GS} = 1.53$.
The differential elastic cross section and, 
in particular, the first dip position and its shape can be compared 
with the data in Fig.\ref{gscal}. The model predictions are given there
by the dashed line. The description is not perfect. 
The general agreement can be improved 
(or rather, the disagreement could be diminished) by a slight change of both
parameters. 
The results for $\lambda =0.1$ and $f_{GS}= 1.64$ 
are presented as thin solid line.

Further improvement can be made by increasing the factor $f_F$ in Eq.(\ref{rho}) 
by the 20\% and respective re-adjustment of $\lambda$ and $f_{GS}$ eventually found to be equal
 to 0.10, and
1.56, respectively. 
The differential elastic cross section for such combination is shown in Fig.\ref{gscal} 
as a thick solid line. The agreement 
with the TOTEM data measured around the first 
diffractive dip position $0.3 < |t| < 0.7$ \cite{totem-dif} is very good.
Comparison with predictions of the other five models presented in Ref.~\cite{totem-dif}
gives a bit of confidence to our simplest proposition of the hadronic interaction picture.

The main point of our interest is, however,
the inelastic (total and elastic) cross sections, which are defined 
mainly with the help of the optical theorem 
by the amplitudes of scattering processes with a very low momentum transfer.

\begin{table}[ht]
\label{tab1}
\begin{center}
\caption{Values of the total, inelastic and elastic cross sections and the 
elastic slope parameter $B$ obtained from different model approximations 
compared with measured values. \label{tab}}
\begin{tabular}{|rl|ccc|c|c|}
\hline 
     &      &  \ \ \ total \ \ \ & \ \ \ inelastic \ \ \ &\ \ \  elastic 
\ \ \ &\ \ \  $B$ \ \ \ & \ \ \ $\sigma_{el}/\sigma_{tot}$ \ \ \ \\
\hline
GS &(Ref.\cite{twds})   &  88.0 &  72.5     & 15.2   & 21.3& 0.17 \\ 
GS &($\lambda$ and $f_{GS}$ adjusted)  &  100.0 & 83.5     & 17.2   & 24.3& 0.17 \\ 
GS &($\lambda$, $f_{GS}$ and $f_F$ adjusted)      &  105.0 & 85.0&  20.0&   27.1&    0.19 \\
\hline 
\multicolumn{2}{|c|}{TOTEM 
\cite{lhc-totem}}
& 98.3 & 73.5\footnotemark[1] & 24.8 & 23.6 & 0.25 \\
\hline
\end{tabular}
\footnotetext[1]{CMS 
 \cite{cms}:
68.0 mb
reported the value of 68.0 mb, ALICE
\cite{alice}:
72.7mb and ATLAS 
\cite{atlas}:
69.4  mb 
for the inelastic cross section.
} 
\end{center}
\end{table}

The value of the inelastic cross section calculated by the respective 
integrations of Eq.(\ref{xsecs})
for $\sqrt{s} = 7$~TeV obtained with the values from Ref.~\cite{twds} of the model
parameters is 72.5 mb and is in surprisingly good agreement with the measurements
made recently not only by the TOTEM Collaboration \cite{lhc-totem}
$(73.5 \pm {{0.6_{\rm stat.}} ^{+1.8}_{{-1.3}^{\rm sys.}}})$ mb,
but also by ATLAS \cite{atlas}
$(69.4 \pm 2.4_{\rm exp.}\pm 6.9_{\rm extr.})$ mb,
CMS 
\cite{cms}
$(68.0 \pm 2.0_{\rm syst.}\pm 2.4_{\rm lum.} \pm 4_{\rm ext.})$ mb
and ALICE \cite{alice}
$(72.7 \pm 1.1_{\rm model}\pm 5.1_{\rm lum.})$ mb.
However, the agreement is worst when comparing the elastic cross section (and
the first dip of the differential elastic cross section shape and position). 
In the first three rows 
of the Table~\ref{tab} we have listed results of the cross section calculations for the model improvements described above. 

Examination of the Table~\ref{tab} leads eventually to the conclusion
that the fraction of the elastic cross section is still too small. 
The combination of $\lambda$, scaling factor $f_{GS}$, and $f_{F}$ modifications 
keeping the first dip of the elastic cross section position and its shape,
as shown above, could not make the situation much better. The 
feature of the dip define two of the model parameters: the first dip 
position fixes the hadronic 
spacial extension - the scale parameter $f_{GS}$ and the value 
of $\lambda$ controls the depth of the dip. 

It is of course possible to obtain the agreement with total, inelastic and elastic cross sections
neglecting the differential elastic data. 
For example, if $f_F = 2 $, $\lambda = 0.2$ and 
$f_{GS} = 1.25$ they are: $\sigma_{tot.} = 98.7$ mb, $\sigma_{inel.} =  74.1$ mb, and 
$\sigma_{el.} = 24.1$ mb.
This fit, however, gives the elastic slope parameter $B$ much less than measured and thus differential elastic cross section values around the first dip are about of a factor of ten higher than they are measured, and, what is even worst, the second dip in the elastic scattering
starts to be visible slightly above $-|t| = 2$ GeV.

\begin{figure}
\centerline{\includegraphics[width=10cm]{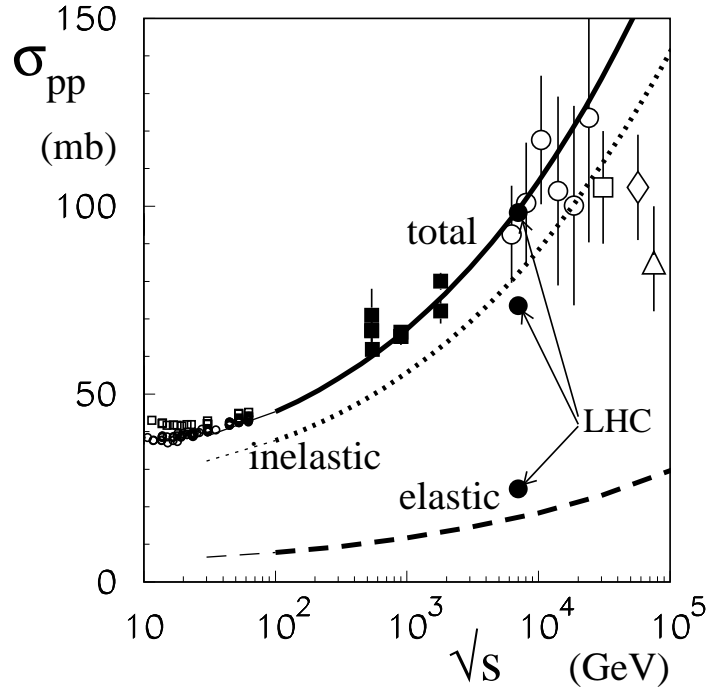}}%
\caption{The energy dependence of $p$-$p$ cross sections
calculated using the $\lambda(s)$, $f_{GS}(s)$ and $f_F(s)$ adjusted using to TOTEM 7 TeV
differential elastic cross section data.
compared with experimental data.
Cosmic ray data rescaled from p-Air to p-p cross section as described in \cite{convert}
are included for comparison: circles from the Akeno \cite{akeno}; square - Fly's Eye \cite{FE}; triangle - Hi-Res \cite{hires}; diamond - PAO \cite{pao}.
}
\label{pps}
\end{figure}

We have thus remain with the minor adjustment $f_F$ value and redefine the energy behaviour of $\lambda(s)$, $f_{GS}(s)$ and $f_F(s)$ with agreement to previous, low energy fits and with the new one at $\sqrt{s}$ of 7 TeV. The final  cross sections are shown in Fig.~\ref{pps}. 
The cosmic ray data for inelastic cross sections \cite{akeno,FE,hires,pao} are shown as an open symbols. They are obtained with the help of the Glauber formalism 
\cite{glauber} described in Ref.\cite{convert}. 
The results of the TOTEM measurement \cite{lhc-totem} are given as solid circles.

\section{Conclusions}
We have shown that the recent differential elastic cross section data in the region of the first dip
could be used to determine the geometrical model parameters at $\sqrt{s} = 7$~TeV,
and the description follow, in general, trends found for lower energy data from 
$ \sqrt{s}\sim 20$~ GeV at FNAL, 53 GeV ISR to SPS with 546 GeV. 

The shape of the first elastic
dip in the differential elastic scattering cross section suggests that the real to imaginary 
part of the elastic amplitude is below the value measured at 
SPS energies and is falling even faster than the COMPETE \cite{compete} prediction.

Comparison of the model extrapolation to the cosmic ray energies with recent shower 
attenuation measurements by Hi-Res and PAO suggests the requirement of the increase of the elastic interaction 
fraction in agreement with the TOTEM measurements.
In the optical picture this could be done by changing of the shape of the opacity function.
This makes the geometrical scaling violated, also the one modified according to the presented way. 
The additional component starts to be visible at present LHC energies. The further increase of the interaction energy will show how fast this component will emerge and how 
much it will contribute to the development of the ultra high-energy cosmic ray cascades
in the atmosphere.

\end{document}